\documentclass[twocolumn]{article} 
\usepackage{epsfig,latexsym}
\begin{document}
\title{Photodetachment study of the $1s3s4s\,^{4}\!S$ resonance in
He$^{-}$}
\author{Andreas E.~Klinkm\"uller, Gunnar Haeffler, Dag Hanstorp, Igor
Yu.~Kiyan\thanks{Permanent address: Russian Academy of Science,
General Physics Institute, Vavilova St.~38, 117\,942 Moscow, Russia
}, Uldis Berzinsh\thanks{Permanent address: Institute of Atomic Physics and
Spectroscopy, University of Latvia, LV 1586 Riga, Latvia},\\ and Christopher
W.~Ingram\thanks{Permanent address: Jesse W.~Beams Laboratory of
Physics, Department of Physics, University of Virginia, Charlottesville,
VA 22\,901, USA}\\
Department of Physics, G\"oteborg University and Chalmers University
of Technology,\\ S-412\,96 G\"oteborg, Sweden
\and
David J. Pegg\\
Department of Physics, University of Tennessee,\\
Knoxville, Tennessee 37\,996, USA
\and
James R.~Peterson\\
Molecular Physics Department, SRI International, \\
Menlo Park, California 94\,025, USA }
\date{23. Jan 1997}
\maketitle
\begin{abstract}
A Feshbach resonance associated with the $1s3s4s\,^{4}\!S$ state of
He$^{-}$ has been observed in the
He$(1s2s\,^{3}\!S)$\,+\,$\mbox{e}^{-}(\epsilon s)$ partial
photodetachment cross section. The residual He($1s2s\,^{3}\!S$) atoms
were resonantly ionized and the resulting He$^{+}$ ions were detected in
the presence of a small background.  A collinear laser-ion beam
apparatus was used to attain both high resolution and sensitivity.  We
measured a resonance energy $E_{\mathrm{r}} = 2.959\,255(7)$ eV and a
width $\Gamma = 0.19(3)$ meV, in agreement with a recent
calculation.
\end{abstract}
\begin{flushleft}
{PACS number(s): 32.80.Fb}
\end{flushleft}
\section{Introduction}
\label{intro}
The metastable He$^{-}$ ion has received considerable attention since
its discovery by Hiby in 1939 \cite{Hib-39}. This simple three
electron ion is the prototype of an unusual class of negative ions that
are not stable, but rather metastable, against autodetachment. The
lowest lying state is the $1s2s2p\,^{4}\!P^{\mathrm{o}}$ state, which is
bound by 77.51(4) meV relative to the $1s2s\,^{3}\!S$ state of He
\cite{Bun-84}. The He$^{-}$ ion in this spin aligned quartet state
cannot radiate and, since it is embedded in a doublet continuum, it can
only autodetach via the relatively weak magnetic interactions. The
varying strengths of these spin dependent interactions result in a
differential metastability among the three fine structure levels. The
longest lived $J=5/2$ level has, for example, a lifetime of 350(15)
$\mu$s \cite{And-93}. Metastable He$^{-}$ ions are therefore
sufficiently long lived to pass through a typical apparatus with only
minor depletion by autodetachment.

Excited states of He$^{-}$, on the other hand, decay rapidly via Coulomb
autodetachment. Their presence is manifested as resonance structure in
scattering cross sections close to thresholds for new channel openings
i.~e.~the excited state energies of the He atom. Many doublet resonances
have been observed, for example, as transient intermediate states in
studies of electron impact on atomic He targets \cite{Buc-94}. Excited
quartet states of He$^{-}$, however, have hitherto received far less
attention. Such states appear as resonances in the photodetachment cross
section and, to a lesser extent, in cross section for detachment via
heavy particle collisions. The energy resolution in
photodetachment measurements is typically much higher then in electron
scattering experiments. This allows one to determine energies and widths
of quartet states more accurately than for corresponding doublet states.
Selection rules on photoexcitation from the
$1s2s2p\,^{4}\!P^{\mathrm{o}}$ ground state of He$^{-}$ allow
transitions only to $^{4}\!S$, $^{4}\!P$ and $^{4}\!D$ excited
states. There have been several experimental \cite{Com-80,Hod-81,Peg-90}
and theoretical studies \cite{Haz-81,Sah-90,Dou-90} of the
photodetachment cross section of He$^{-}$. Most recently, Xi and Froese
Fischer \cite{Xi-96} have calculated the position and width of quartet
states of He$^{-}$ below the He($n=4$) thresholds.

The prominent $1s2p^{2}\,^{4}\!P$ shape resonance, which lies just above
the He($1s2p\,^{3}\!P^{\mathrm{o}}$) threshold, was first observed
experimentally by Peterson et al \cite{Pet-85} in a laser
photodetachment study. The accuracy of the measured resonance parameters
was later improved by Walter et al \cite{Wal-94-4}.

In the present paper we report on the first investigation of the doubly
excited $1s3s4s\,^{4}\!S$ state in He$^{-}$, which appears as a Feshbach
resonance in the He$(1s2s\,^{3}\!S)$\,+\,$\mbox{e}^{-}(\epsilon s)$ partial
cross section just below the He($1s3s\,^{3}\!S$) threshold. 
\section{Experiment}
\label{expe}
\subsection{Procedure}
\begin{figure}
\begin{center}
\epsfig{file=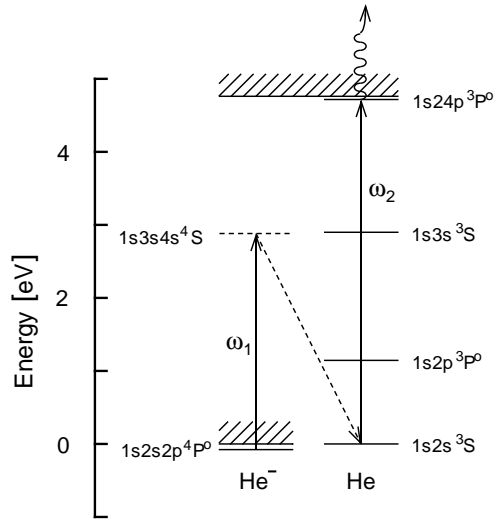, width=0.8\columnwidth}
\end{center}
\caption{Excitation scheme showing selected states of He/He$^{-}$.  The
solid arrows represent the transitions induced in this experiment. The
dashed arrow indicates one autodetachment channel of the doubly excited
state of He$^{-}$. The wavy arrow indicates field ionization.}
\label{term}
\end{figure}

A two-color excitation scheme (Fig.\ \ref{term}) was used to isolate the
He$(1s2s\,^{3}\!S)+\mbox{e}^{-}(\epsilon s, \epsilon d)$ partial
photodetachment cross sections in this experiment.  The procedure was
previously developed for studies of threshold behavior \cite{Hae-96-1}
and resonance structure \cite{Lju-96} in the photodetachment cross
section of Li$^{-}$.  In the first step, metastable He$^{-}$ ions were
detached by absorption of photons of frequency $\omega_{1}$.  In the
second step He atoms, left in the $1s2s\,^{3}\!S$ state as result of the
photodetachment, were resonantly excited to a high lying Rydberg state
by absorption of photons of frequency $\omega_{2}$ and subsequently
field ionized in a static electric field of about 150 kV/m.  These
processes can be described by the following equations:
\begin{eqnarray}\label{exec}
\nonumber\mbox{He}^{-} (1s2s2p\,^{4}\!P^{\mathrm{o}}) + \hbar \omega_{1}
& \!\!\!\!\rightarrow\!\!\!\! & \mbox{He} (1s2s\,^{3}\!S) +
\mbox{e}^{-}(\epsilon s,\epsilon d), \\ \label{exe2} \mbox{He}
(1s2s\,^{3}\!S) + \hbar \omega_{2} & \!\!\!\!\rightarrow\!\!\!\! &
\mbox{He} (1s24p\,^{3}\!P^{\mathrm{o}}), \\
\label{exe3} \mbox{He} (1s24p\, ^{3}\!P^{\mathrm{o}}) &
\!\!\!\!\leadsto\!\!\!\! &
\mbox{He}^{+} (1s\, ^{2}\!S) + \mbox{e}^{-}
\end{eqnarray}
where $\leadsto$ represents field ionization.  

The yield of the He$^{+}$ ions produced in this state selective
detection scheme was recorded as a function of the frequency
$\omega_{1}$, while the frequency $\omega_{2}$ was held constant on the
transition to the Rydberg state. Thus, only photodetachment into the
He$(1s2s\,^{3}\!S_{1})$\,+\,$\mbox{e}^{-}(\epsilon s, \epsilon d)$
channels contributed to the He$^{+}$ signal. Since both laser
intensities were constant during a scan, the positive ion signal, as
function of $\omega_{1}$, is proportional to the
He$(1s2s\,^{3}\!S_{1})$\,+\,$\mbox{e}^{-}(\epsilon s, \epsilon d)$
partial photodetachment cross sections.

The detection scheme based on the selective detection of residual
He($1s2s\,^{3}\!S$) atoms was effective in eliminating a potential
background source unique to measurements involving metastable negative
ions. Since He$^{-}$ ions are non-stable, the ion beam will contain a
fraction of He atoms produced, in flight, by autodetachment. These He
atoms will, however, be in the $1s^{2}\,^{1}\!S$ ground state and
therefore will not be resonantly photoionized.

\subsection{Experimental arrangement}
The $^{4}$He$^{-}$ beam was produced from a mass selected He$^{+}$ ion
beam via charge exchange in a Cs vapor cell.  The beam energy was 3.1
keV. A current of typically 1 nA was obtained in the interaction region.

\begin{figure}
\begin{center}
\epsfig{file=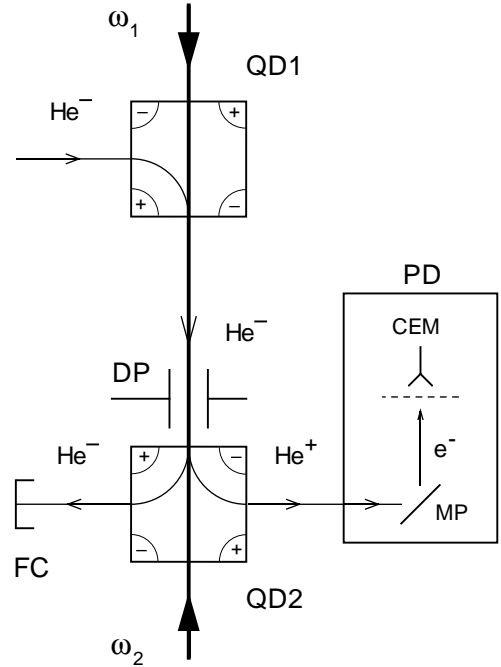, width=0.8\columnwidth}
\end{center}
\caption{Interaction-detection region: {\sf QD1,QD2,} {electrostatic
quadrupole deflectors}; {\sf CEM,} {channel electron multiplier}; {\sf
DP,} {deflection plates}; {\sf PD,} {positive ion detector}; {\sf FC,}
{Faraday cup}; {\sf MP,} {metal plate}. Ion- and laser beams were merged
in the 0.5~m long interaction region between the quadrupole deflectors.}
\label{exp}
\end{figure}

In the interaction region shown schematically in Fig.\
\protect\ref{exp}, the laser and ion beams were coaxially superimposed
over the 0.5~m path between the two electric quadrupole deflectors ({\sf
QD1, QD2}).  The beam paths were defined by apertures of 3~mm diameter
at both ends of the interaction region. The apparatus has been
previously described in more detail \cite{Han-95}.

The apparatus was designed to reduce the background of He$^{+}$ ions
produced by double collisional detachment by installing a pair of
deflection plates ({\sf DP}) just before the second quadrupole deflector
({\sf QD2}).  The transverse electric field between the deflection
plates was insufficient to field ionize the $24p$ Rydberg atoms, but
strong enough to sweep collisionally created He$^{+}$ ions out of the
beam.

The highly excited He atoms were field ionized by the static electric
field in the second quadrupole deflector ({\sf QD2}). The resulting
He$^{+}$ ions were deflected into the positive ion detector ({\sf PD}),
where they impinged on a metal plate ({\sf MP}) and produced secondary
electrons that were detected with a channel electron multiplier ({\sf
CEM}).  To monitor the He$^{-}$ beam current in the Faraday cup ({\sf
FC}) we periodically grounded the deflection plates.

The two photon frequencies $\omega_{1}$ and $\omega_{2}$ used in the
experiment were produced by a pair of dye lasers pumped by a common XeCl
excimer laser that delivered pulses of about 15 ns duration.  The
visible laser light of frequency $\omega_{1}$ was generated by use of an
Exalite 416 dye. The energy in this case was about 1~mJ. The UV laser
light of frequency $\omega_{2}$ was obtained by frequency doubling the
fundamental output generated by a Coumarin 334 dye. In this case the
pulse energy was typically 0.05~mJ. The output of laser $\omega_{1}$ was
attenuated to avoid saturation of both the photodetachment process and
the detection system. The two laser pulses counter-propagated and
overlapped in the interaction region.

\begin{table}
\protect\caption{Argon calibration lines \protect\cite{Min-73}: The
levels are designated in Paschen notation.  For counter-propagating
laser ($\omega_{1}$) and ion beams we used the lines number 1, 2 and 3
and for co-propagating laser ($\omega_{1}$) and ion beam we used the
lines number 2, 3 and 4.}\medskip
\label{Argon}
\begin{center}
\begin{tabular}{lcc}
\hline\hline
\# & Transition &  Wavenumber (cm$^{-1}$) \\
\hline
1 & 3p$_{5}$\,$\rightarrow$\, 1s$_{4}$ & 23\,812.359(2) \\
2 & 3p$_{4}$\,$\rightarrow$\, 1s$_{3}$ & 23\,853.767(2) \\
3 & 3p$_{8}$\,$\rightarrow$\, 1s$_{5}$ & 23\,855.565(2) \\
4 & 3p$_{2}$\,$\rightarrow$\, 1s$_{3}$ & 23\,905.933(2) \\
\hline
\end{tabular}
\end{center}
\end{table}

The frequency $\omega_{1}$ was determined by combining Fabry-Perot
fringes with optogalvanic spectroscopy. The Fabry-Perot fringes served
as frequency markers whereas the four transitions in Ar
given in Table \protect\ref{Argon}, generated in a hollow
cathode lamp, provided an absolute calibration of the energy
scale. The energy resolution of this apparatus has been
demonstrated to be approximately 0.2~cm$^{-1}$ \cite{Hae-96-1}, which
is the linewidth of the laser.

\section{Results and discussion}
\label{con}
\begin{figure}
\epsfig{file=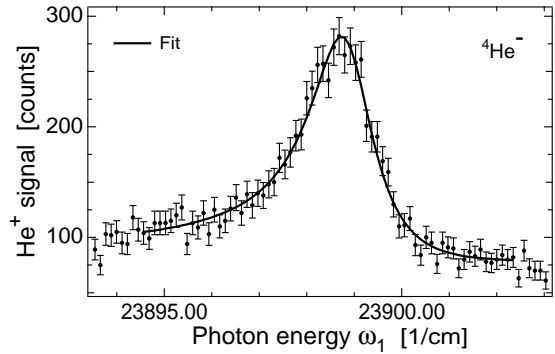,width=\columnwidth}
\caption{He$^{-}(1s3s4s\,^{4}\!S)$ resonance: Measurement of the
He$(1s2s\,^{3}\!S)+\mbox{e}^{-}(\epsilon s, \epsilon d)$ partial
photodetachment cross sections of He$^{-}$ in the vicinity of the
$1s3s4s\,^{4}\!S$ resonance. The use of co-propagating laser
($\omega_{1}$) and ion beams causes the resonance to be Doppler shifted
by 30.89 cm$^{-1}$ to the blue. The solid line is a fit to the data
using Eq.\ (\protect\ref{Sho}) (for details, see text).  The error bars
represent the shot noise.  Each data point represents 200 laser pulses.}
\label{res}
\end{figure}

A typical measurement of the $1s3s4s\,^{4}\!S$ resonance is shown in
Fig.~\ref{res}. The resonance structure appears on a background created
by several processes. We found three processes to significantly
contribute to this background: Firstly (a), collisional detachment
leaving the He atom in the $1s2s\,^{3}\!S$ state. Secondly (b),
photodetachment by the laser $\omega_{2}$
\begin{displaymath}
\mbox{He}^{-} (1s2s2p\, ^{4}\!P^{\mathrm{o}}) + \hbar
\omega_{2} \rightarrow  \mbox{He} (1s2s\, ^{3}\!S) +
\mbox{e}^{-}(\epsilon s,\epsilon d),
\end{displaymath}
followed by resonance ionization indicated in Eq.\ (\protect\ref{exe2})
and Eq.\ (\protect\ref{exe3}). Thirdly (c), non-resonant photodetachment
by the laser of frequency $\omega_{1}$
\begin{equation}\label{exec2}
\mbox{He}^{-} (1s2s2p\,^{4}\!P^{\mathrm{o}}) + \hbar
\omega_{1} \rightarrow  \mbox{He} (1s2s\,^{3}\!S) +
\mbox{e}^{-}(\epsilon d),  
\end{equation}
and then proceeding as in Eq.\ (\protect\ref{exe2}) and Eq.\
(\protect\ref{exe3}).

To reduce the collisional detached contribution (a) we maintained a
pressure of $5\times 10^{-9}$ mbar ($5\times 10^{-7}$ Pa) in the
interaction chamber. To reduce contribution (b) we attenuated the
output of laser $\omega_{2}$.

The selection rule $\Delta L=0$ for Coulomb autodetachment in
$LS$-coupling forbids the $1s3s4s\,^{4}\!S$ resonance from appearing in
the He$(1s2s\,^{3}\!S)$\,+\,$\mbox{e}^{-}(\epsilon d)$ partial cross
section. Thus, the process (c), as represented by Eq.\
(\protect\ref{exec2}), contributes a constant background over the region
shown in Fig.\ \protect\ref{res}.  Xi and Froese Fischer \cite{Xi-96-2}
predict a $d$-wave photodetachment cross section of 6 Mb across the
resonance and a peak $s$-wave cross section of approximately 20 Mb.

The cross section $\sigma_{\mathrm{Sh}}(E)$ around the
resonance can be parametrized according to Shore \cite{Sho-68},
\begin{eqnarray}\label{Sho}
\sigma_{\mathrm{Sh}}(E) &=& a+\frac{b\epsilon+c}{1+\epsilon^{2}}\\
\epsilon &=& \frac{E-E_{0}}{(\Gamma /2)}\quad ,\nonumber
\end{eqnarray}
where $E_{0}$ is the resonance energy, $\Gamma$ the resonance width, $E$
the photon energy, $a$ the background cross section, and $b,c$ are the
Shore parameters. This function is least-square fitted to our positive
ion signal. The Shore parametrization method applies to either total or
partial photodetachment cross sections.

The data shown in Fig.~\ref{res} are recorded with co-propagating laser
($\omega_{1}$) and ion beams. The fit yields a value for the
blue-shifted resonance energy $E_{0}^{\mathrm{b}}$. To eliminate the
Doppler shift, we repeated the measurements using counter-propagating
laser and ion beams to determine the red-shifted resonance energy,
$E_{0}^{\mathrm{r}}$. The resonance energy, corrected for the Doppler
effect to all orders, $E_{\mathrm{r}}$, is given by the geometric mean
of the two measurements:
\begin{eqnarray*}
E_{\mathrm{r}} &=& \sqrt{E_{0}^{\mathrm{b}} E_{0}^{\mathrm{r}}}\\ &=&
23\,867.992(55)\;\mbox{cm$^{-1}$}\quad .\nonumber
\end{eqnarray*}

This result is an average obtained from 7 spectra taken with
co-propagating laser and ion beams and 8 spectra taken with
counter-propagating beams. There are two major contributions to the
quoted uncertainty: 0.01~cm$^{-1}$ is due to the calibration
uncertainty, and the remainder is due to statistical scatter of the
fitted resonance parameters. In Table \protect\ref{tab1} the values are
compared with recently calculated resonance parameters.

The measured resonance parameters agree with those calculated by Xi and
Froese Fischer \cite{Xi-96} within the limited precision of the
latter. Our measurement is, however, almost two orders of magnitudes
more precise and it should stimulate further theoretical work.
\begin{table}
\protect\caption{Comparison of the present measurement and a calculation
of the $1s3s4s\,^{4}\!S$ resonance in He$^{-}$. The recommended
conversion factor of (1/8\,065.5410)
$\big[\mbox{eV}/(\mbox{cm$^{-1}$})\big]$ was used
\protect\cite{Coh-88}.}\medskip
\label{tab1}
\begin{center}
\begin{tabular}{lr@{.}lr@{.}l}
\hline\hline Author & \multicolumn{2}{c}{$E_{\mathrm{r}}$ (eV)} &
\multicolumn{2}{c}{$\Gamma$ (meV)} \\ \hline {\it
Theory}:&\multicolumn{2}{c}{}&\multicolumn{2}{c}{}\\ Xi et al
\cite{Xi-96} (1996)&\multicolumn{2}{c}{}&\multicolumn{2}{c}{} \\ Length
form & 2&959\,07 & 0&19 \\ Velocity form & 2&959\,08 & 0&18 \\ {\it
Experiment}:&\multicolumn{2}{c}{}&\multicolumn{2}{c}{}\\ This work &
2&959\,255(7) & 0&19(3) \\ \hline
\end{tabular}
\end{center}
\end{table}
\section{Summary}
\label{summary}
In the present experiment we have combined laser photodetachment and
resonance ionization to study a Feshbach resonance associated with the
autodetaching decay of the double excited $1s3s4s\,^{4}\!S$ state in
He$^{-}$. The measurement was made under the simultaneous conditions of
high sensitivity and energy resolution using a collinear beam
apparatus. The production of doubly excited states was enhanced by the
use of a large interaction volume defined by the overlap of the
co-axially superposed ion and laser beams. The residual atoms produced
in the photodetachment process were selectively detected according to
their excitation state to reduce the background. The energy resolution
is enhanced in a collinear beam measurement by kinematic compression of
the longitudinal velocity distribution of the fast moving ions
\cite{Kau-76}. In addition, the Doppler shift was removed, to \emph{all}
orders, by performing separate measurements using co- and
counter-propagating laser and ion beams.

The frequency doubled output of excimer pumped dye lasers can be used,
in principle, to study the photodetachment cross section of He$^{-}$ up
to the double detachment limit. Moreover, the decay of a given doubly
excited state can be studied in different channels.  We plan to search
for resonances below higher lying thresholds and to investigate
different decay channels.

\section{Acknowledgement}
Xi and Froese Fischer are acknowledged for having provided us with
unpublished data.  Financial support for this research has been obtained
from the Swedish Natural Science Research Council (NFR). Personal
support was received from Wenner-Gren Center Foundation for Igor Kiyan
and from Nordisk Forskerutdanningsakademi (NorFA) for James Peterson.
David Pegg acknowledges the support from the Swedish Institute and the
U.S.~Department of Energy, Office of Basic Energy Sciences, Division of
Chemical Sciences.
%
%

\begin{thebibliography}{10}

\bibitem{Hib-39}
Julius~W. {Hiby}.
\newblock Massenspektrographische {U}ntersuchungen an {W}asserstoff- und
  {H}eliumstrahlen ({H}$^{+}_{3}$, {H}$^{-}_{2}$, {H}e{H}$^{+}$,
{H}e{D}$^{+}$,
  {H}e$^{-}$).
\newblock {\em Analen der Phys.}, 34:473--487, 1939.
\newblock (5. Folge).

\bibitem{Bun-84}
Annik~Vivier {Bunge} and Carlos~F. {Bunge}.
\newblock Improved calculation of the electron affinity of {H}e
  $1s2s\,^{3}\!{S}$.
\newblock {\em Phys. Rev.~A}, 30(5):2179--2182, November 1984.

\bibitem{And-93}
T.~{Andersen}, L.~H. {Andersen}, P.~{Balling}, H.~K. {Haugen},
P.~{Hvelplund},
  W.~W. {Smith}, and K.~{Taulbjerg}.
\newblock Metastable-ion lifetime studies utilizing a heavy-ion storage
ring:
  Measurements on {H}e$^{-}$.
\newblock {\em Phys. Rev.~A}, 47(2):890--896, February 1993.

\bibitem{Buc-94}
Stephen~J. {Buckman} and Charles~W. {Clark}.
\newblock Atomic negative-ion resonances.
\newblock {\em Rev. Mod. Phys.}, 66(2):539--655, April 1994.

\bibitem{Com-80}
R.~N. {Compton}, G.~D. {Alton}, and D.~J. {Pegg}.
\newblock Photodetachment cross sections for
  {H}e$^{-}(^{4}\!{P}^{\mathrm{o}})$.
\newblock {\em J. of Phys.~B: At., Mol. Opt.}, 13:L651--L655, 1980.

\bibitem{Hod-81}
R.~V. {Hodges}, M.~J. {Coggiola}, and J.~R. {Peterson}.
\newblock Photodetachment cross sections for {H}e$^{-}$ $^{4}\!{P}$.
\newblock {\em Phys. Rev.~A}, 23(1):59--63, January 1981.

\bibitem{Peg-90}
D.~J. {Pegg}, J.~S. {Thompson}, J.~{Dellwo}, R.~N. {Compton}, and G.~D.
  {Alton}.
\newblock Partial cross sections for the photodetachment of metastable
  {H}e$^{-}$.
\newblock {\em Phys. Rev. Lett.}, 64:278--281, 1990.

\bibitem{Haz-81}
A.~U. {Hazi} and K.~{Reed}.
\newblock Theoretical photodetachment cross section for
  {H}e$^{-}$($^{4}\!{P}^{\mathrm{o}}$).
\newblock {\em Phys. Rev.~A}, 24(4):2269--2272, October 1981.

\bibitem{Sah-90}
H.~P. {Saha} and R.~N. {Compton}.
\newblock Theoretical studies of the photophysics of
  {H}e$^{-}$($1s2s2p$)$^{4}\!{P}^{\mathrm{o}}$.
\newblock {\em Phys. Rev. Lett.}, 64(13):1510--1513, March 1990.

\bibitem{Dou-90}
Maryvonne~Le {Dourneuf} and Shinichi {Watanabe}.
\newblock Grandparent model of the doubly excited {H}e$^{-\ast\ast}$
resonances  from a hypersperical viewpoint.
\newblock {\em J. of Phys.~B: At., Mol. Opt.}, 23:3205--3224, 1990.

\bibitem{Xi-96}
Jinhua {Xi} and Charlotte {Froese Fischer}.
\newblock Cross section and angular distribution for the photodetachment of
  {H}e$^{-}$($1s2s2p\,^{4}\!{P}^{\mathrm{o}}$) below the {H}e($n=4$)
threshold.
\newblock {\em Phys. Rev.~A}, 53(5):3169--3177, May 1996.

\bibitem{Pet-85}
J.~R. {Peterson}, Y.~K. {Bae}, and D.~L. {Huestis}.
\newblock Threshold behavior near an electronic shape resonance:
Analysis of
  the $\mbox{{H}e}(^{3}\!{P})$ threshold in $\mbox{{H}e}^{-}$
photodetachment
  and determination of the $\mbox{{H}e}(2^{3}\!{S})$ electron affinity.
\newblock {\em Phys. Rev. Lett.}, 55(7):692--695, August 1985.
\bibitem{Wal-94-4}
C.~W. {Walter}, J.~A. {Seifert}, and J.~R. {Peterson}.
\newblock Reexamination of the {H}e$^{-}$
$1s2p^{2}\,^{4}\!{P}^{\mathrm{e}}$
  shape resonance: Details of its properties and a precise electron
affinifty for {H}e $2\;^{3}\!{S}$.
\newblock {\em Phys. Rev.~A}, 50(3):2257--2262, September 1994.

\bibitem{Hae-96-1}
Gunnar {Haeffler}, Dag {Hanstorp}, Igor {Kiyan},
Andreas~E. {Klinkm{\"u}ller},
  and Ulric {Ljungblad}.
\newblock Electron affinity of {L}i: {A} state-selective measurement.
\newblock {\em Phys. Rev.~A}, 53(6):4127--4131, June 1996.
\newblock e-print: physics/9703013.

\bibitem{Lju-96}
U.~{Ljungblad}, D.~{Hanstorp}, U.~{Berzinsh}, and D.~J. {Pegg}.
\newblock Observation of doubly excited states in {L}i$^{-}$.
\newblock {\em Phys. Rev. Lett.}, 77(18):3751--3754, October 1996.

\bibitem{Han-95}
Dag {Hanstorp}.
\newblock An ion beam apparatus for collinear photodetachment experiments.
\newblock {\em Nucl. Instrum. Methods B}, 100:165--175, 1995.

\bibitem{Min-73}
Lennart {Minnhagen}.
\newblock Spectrum and the energy levels of neutral argon, {A}r {\sc i}.
\newblock {\em J. Opt. Soc. Am.}, 63(10):1185--1198, 1973.

\bibitem{Xi-96-2}
Jinhua {Xi} and Charlotte {Froese Fischer}.
\newblock private communication, 1996.

\bibitem{Sho-68}
Bruce~W. {Shore}.
\newblock Parametrization of absorption line profiles.
\newblock {\em Phys. Rev.}, 171(1):43--54, July 1968.

\bibitem{Coh-88}
E.~Richard {Cohen} and Barry~N. {Taylor}.
\newblock The 1986 {CODATA} recommended values of the fundamental physical
  constants.
\newblock {\em J. Phys. Chem. Ref. Data}, 17(4):1795--1803, 1988.

\bibitem{Kau-76}
S.~L. {Kaufman}.
\newblock High-resolution laser spectroscopy in fast beams.
\newblock {\em Opt. Comm.}, 17(3):309--312, 1976.

\end{thebibliography}
%
%

%
\end{document}